\documentclass{blois}

\def\ra{\rightarrow}

\def\be{\begin{equation}}
\def\ee{\end{equation}}
\def\bea{\begin{eqnarray}}
\def\eea{\end{eqnarray}}

\newcommand{\Photo}{}

\begin{document}
\begin{flushright}
 LPSC15250
\end{flushright}

\vspace*{4cm}
\title{Revisiting LHC gluino mass bounds through radiative decays using MadAnalysis 5}

\author{G. CHALONS \footnote{Speaker}, D. SENGUPTA}

\address{Laboratoire Physique Subatomique et Cosmologie, Universit\'{e} Grenoble-Alpes,\\
 CNRS/IN2P3, 53 rue des Martyrs, F-38026 Grenoble, France}

\maketitle\abstracts{The ATLAS and CMS experiments at the CERN LHC have collected about 25
$\mbox{fb}^{-1}$ of data each at the end of their 8 TeV run, and ruled out a huge swath of parameter space
in the context of Minimally Supersymmetric Standard Model (MSSM). Limits on masses
of the gluino ($\widetilde g$) have been pushed to above 1 TeV. These limits are however extremely
model dependent and do not always reflect the level of exclusion. So far the limits on
the gluino mass using the simplified model approach only constrained its value using its
three-body decays. We show in this work that already existing ATLAS and CMS analysis
can also constrain the radiative gluino decay mode and we derived improved mass limits
in particular when the mass difference between the LSP and the gluino is small.}

\section{Introduction and Motivation}
Supersymmetry (SUSY) is one of the best motivated and most studied beyond Standard Model (BSM) paradigm. SUSY has an extremely rich phenomenology since it predicts a lot of new particles which could lie around the electroweak scale, which can be searched for at the LHC~\cite{atlas-susy-twiki,cms-susy-twiki}. In particular, the superpartners of the gluons, the gluinos $\widetilde g$, possess the largest pair-production cross section at the LHC and are therefore intensively searched for by experimental collaborations. Since no signal of SUSY has been uncovered, the only appropriate interpretation of these null searches was to set limits on the production cross section and masses of superpartners. However, interpreting the searches for NP is a non-trivial task and almost impossible to perform in a model independent way. The first focus of SUSY searches was to set limits on constrained Minimal Supersymmetric Standard Model (CMSSM) or minimal SUperGRAvity (mSUGRA) scenarios~\cite{atlas-susy-twiki,cms-susy-twiki}. The current lower limits on the gluino stand at, in the CMSSM, $m_{\tilde g} < 1.7$ TeV for almost degenerate gluino and squarks. Nevertheless these limits are extremely model dependent and do not cover all possibilities in which a SUSY signal could show up. Indeed, if the lightest SUSY particle (LSP) is massive and degenerate with the squarks and/or gluinos, the so-called ``compressed SUSY'' scenarios, these limits can be seriously weakened. To relax some of these assumptions, the ATLAS and CMS collaborations have adopted the Simplified Model Spectra (SMS) approach (see for example~\cite{atlas-susy-twiki,cms-susy-twiki,Alves:2011wf} and references therein) to interpret the NP searches in a less model-dependent way. 

Whether from the CMSSM/mSUGRA or the SMS interpretations of SUSY searches, if the first and second generation squarks are degenerate, the limits on their masses are quite strong and are pushed above the TeV scale. In this case, with light flavour squarks decoupled $m_{\tilde g} \gg m_{\tilde q}$, then the two body decays $\widetilde g \rightarrow q \widetilde q$ are forbidden. In this situation, as far as the official experimental analyses are concerned, published gluino mass limits in the SMS approach focussed on its three-body decays $\widetilde g \rightarrow q \bar{q} \widetilde{\chi}_1^0$ or $\widetilde g \rightarrow q q' \widetilde{\chi}_1^\pm$~\cite{atlas-susy-twiki,cms-susy-twiki}, where $\widetilde{\chi}_1^0,\widetilde{\chi}_1^\pm$ are the lightest neutralino and chargino respectively. These decays are mediated by the squark corresponding to the flavour of the final states quarks, in the case where the couplings of the gluinos are flavour-symmetric. 

Nevertheless, since the limits on the first and second generation pushed their masses beyond the TeV scale, the three-body decays may be heavily suppressed. Moreover, even in the case where the third generation squarks are allowed to be lighter, if the mass difference $\Delta M = m_{\tilde g}-m_{{\widetilde{\chi}}_1^0}$ is small, then the three-body decays are kinematically suppressed and official analyses loose their sensitivity. However, there exists one decay mode which does not suffer from such suppression factors, which had received little attention so far: that of the two-body radiative decay $\widetilde g \rightarrow g \widetilde{\chi}_1^0$.

\section{The radiative gluino decay} 
\begin{figure}[t]
\begin{center}

\mbox{
\includegraphics[width=.49\textwidth]{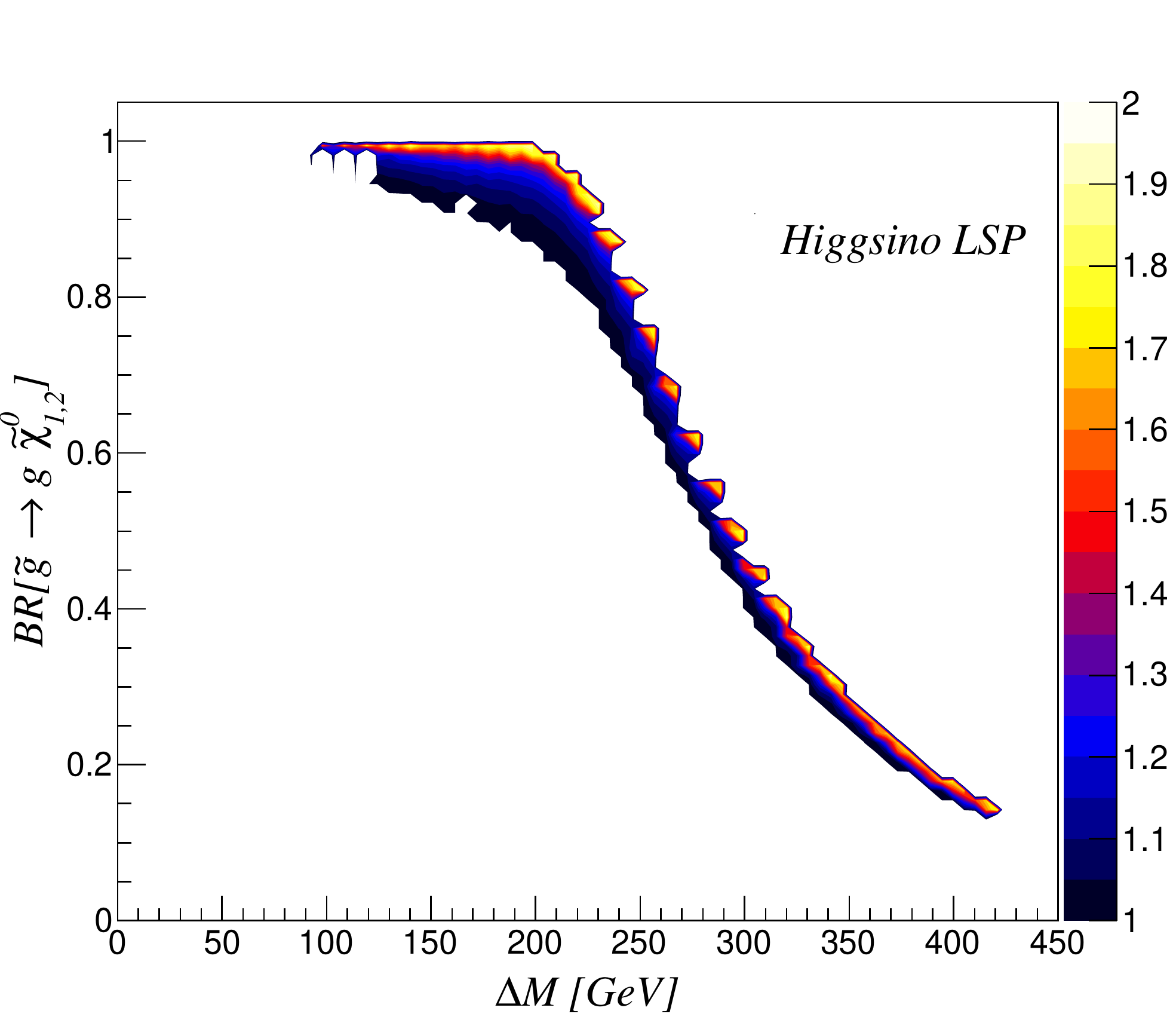}
\includegraphics[width=.49\textwidth]{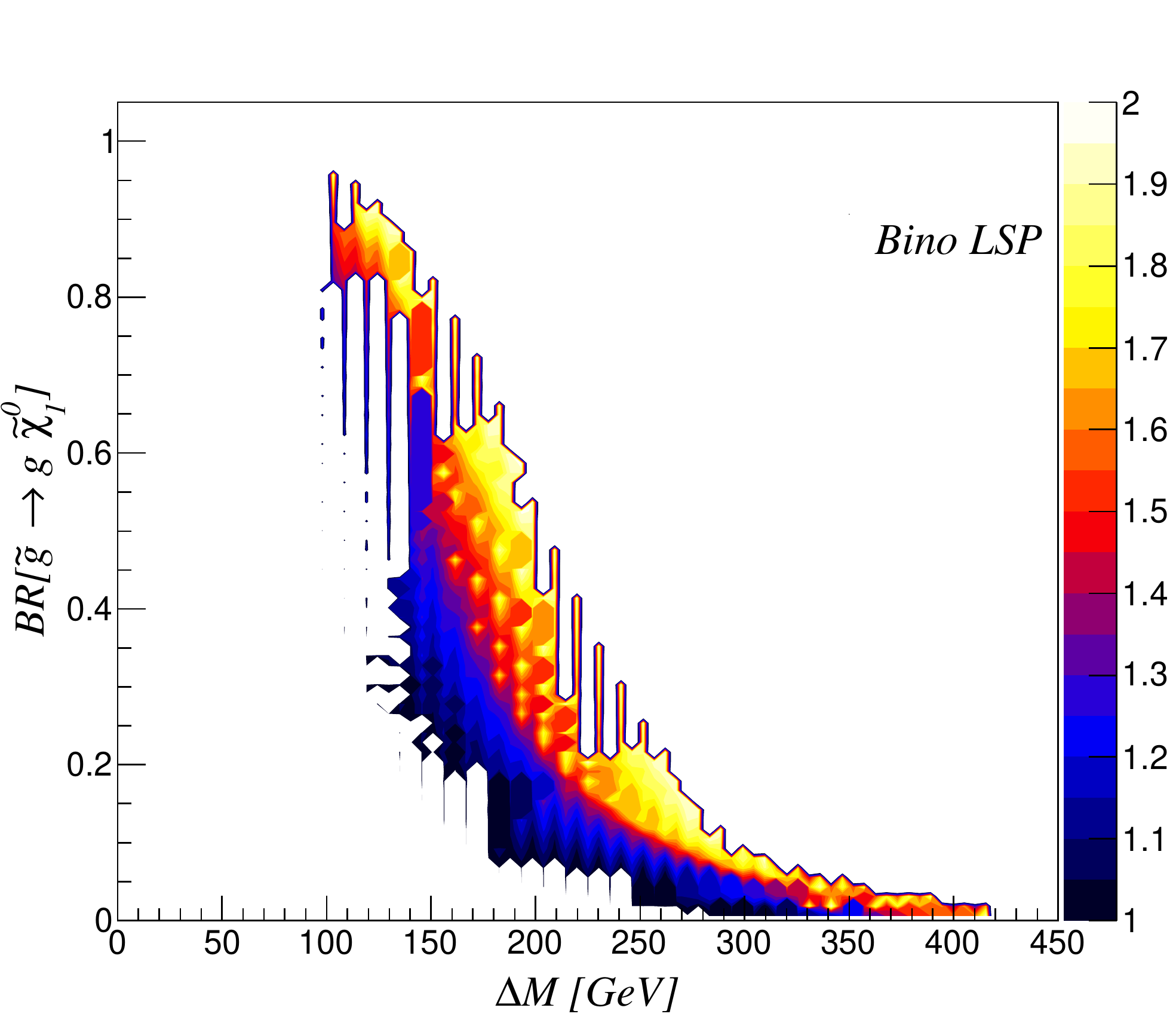}}

\end{center}
\caption{Branching ratio of $\widetilde g \ra g \widetilde{\chi}_1^0$ 
in terms of $\Delta M = m_{\tilde g}-m_{{\widetilde{\chi}}_1^0}$ for a higgsino-like (left panel) and bino-like 
(right panel) neutralino. In the higgsino-like case the branching ratio of the radiative decay $\widetilde g \rightarrow g \widetilde{\chi}_2^0$ is also included since the decay products of $\widetilde{\chi}_2^0$ are very soft as $m_{{\widetilde{\chi}}_2^0}-m_{{\widetilde{\chi}}_1^0}$ is small. The colour shading indicates the ratio $m_{{\tilde b}_1}/m_{{\tilde t}_1}$.}
\label{fig:bratios}
\end{figure}

The gluino radiative decay $\widetilde g \rightarrow g \widetilde{\chi}_1^0$ is induced dominantly by stops/tops loops. The neutralino $\widetilde{\chi}_1^0$ is an admixture of wino- ($\widetilde{W}_3$), bino-
($\widetilde B$) and higgsino-like ($\widetilde H$) neutral spinors, which are 
respectively the superpartners of the neutral gauge eigenstates $B,W_3$  and of the two 
Higgs doublet of the MSSM $H_{u,d}$.

From an effective Lagrangian point of view, where all the squarks are integrated out, the decay width into a wino-like neutralino is strongly suppressed since it is induced by a dimension seven operator~\cite{Gambino:2005eh}. For a bino-like or higgsino-like neutralino, the loop decay is induced by a dimension five chromo-magnetic operator,
\begin{equation}
\label{dim5gluino}
 {\cal L}_{\rm eff.} = \frac{1}{\widetilde m}\overline{\widetilde{\chi}_1^0} \sigma^{\mu\nu}
P_{L,R} \widetilde{g}^a G_{\mu\nu}^b \delta_{ab}
\end{equation}
where $\widetilde m$ is an effective squark mass scale. In the Higgsino case, this operator is obtained in the effective theory from a top-top-gluino-higgsino operator in which the two top quarks form a loop which emits a gluon. Such a diagram is divergent in the effective theory thereby generating a logarithmic enhancement $m_t^2/m_{\tilde g}^2\ln\left(m_{\tilde t}^2/m_t^2\right)$, where $m_{\tilde t}$ is a common top squark mass, see~\cite{Gambino:2005eh} and references therein. The radiative gluino decay into a bino-like neutralino does not benefit from such enhancement. In a scenario with very heavy squarks, where both the radiative and three-body modes are available for the gluino to decay into, the logarithmic enhancement can be such that the loop decay dominates over other branching fractions. In case of very large squark masses, large corrections originating from the large logarithm should be resummed~\cite{Gambino:2005eh}. If the third generation squarks are somewhat lighter than the the two other generations, and the $t \bar{t}$ threshold is closed, the three-body decay $\widetilde g \rightarrow b \bar{b} \widetilde{\chi}_1^0$ can still compete with the radiative decay. In this case a hierarchical third generation squark spectrum, with sbottoms heavier than the stops, can suppress the three-body decay into $b$-quarks with respect to the loop decay, which can then dominate, see Fig.~\ref{fig:bratios}, taken from~\cite{Chalons:2015vja}.

\section{Gluino mass limits using the radiative decay}
\begin{figure}[t]
 \begin{center}
  \includegraphics[width=.49\textwidth]{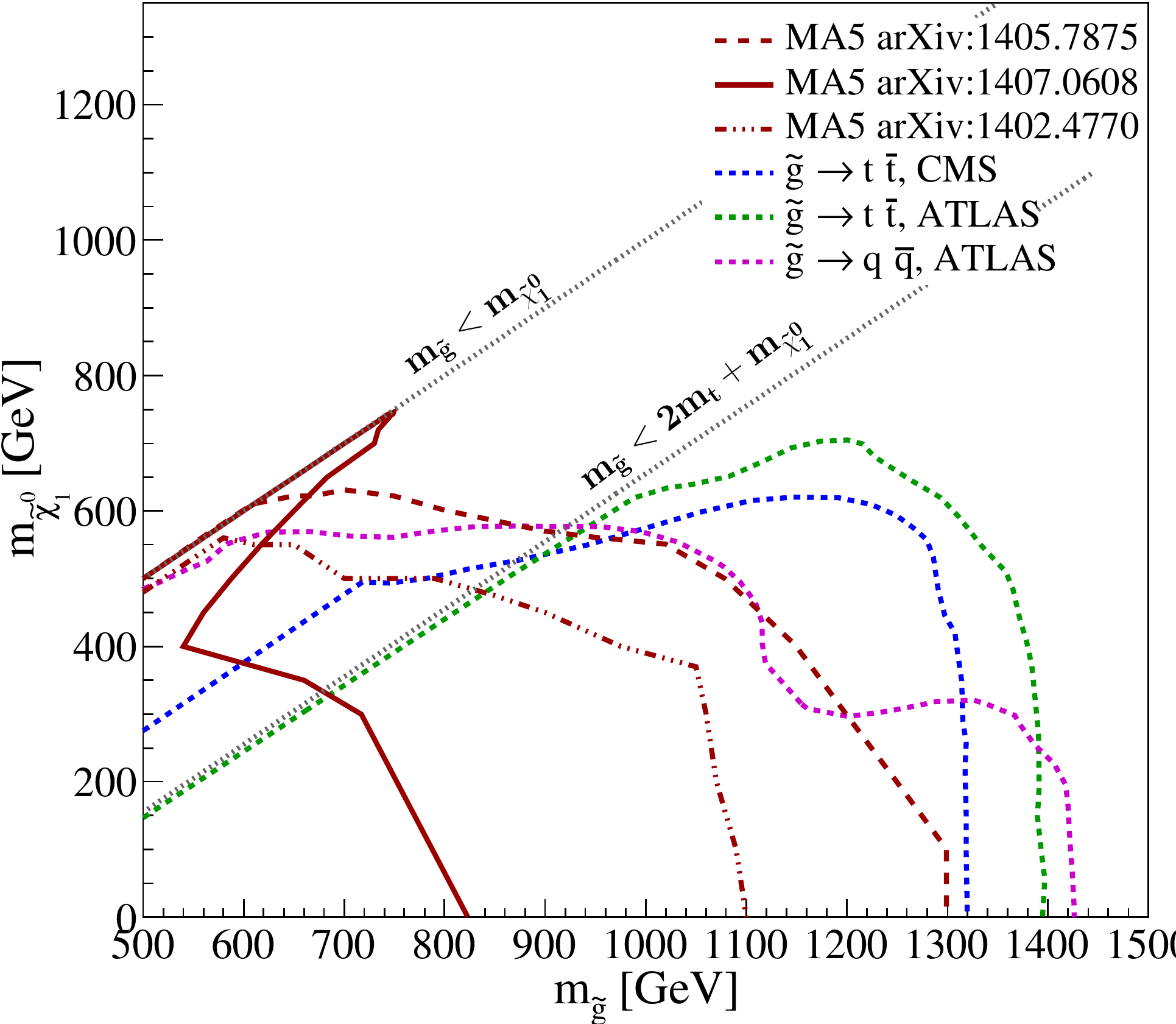}
 \end{center}\caption{\label{fig:8TeVlimits}95\% CL exclusion 
contours for the radiative gluino decay simplified topology. The solid red line 
corresponds to the mass limits obtained from the \texttt{MA5} recasted
ATLAS monojet analysis, the red broken line from the \texttt{MA5} 
recasted ATLAS multijet search and the dashed-dotted line from the 
\texttt{MA5} recasted CMS multijet analysis.
For comparison, the official 95\% CL exclusion lines for the $\widetilde{g} \ra t \bar{t} \widetilde{\chi}_1^0$ SMS from ATLAS (green dashed line)and CMS (blue dashed line) and for $\widetilde{g} \ra q \bar{q} \widetilde{\chi}_1^0$ SMS from ATLAS (purple dashed line) are also shown. 
}
\end{figure}

Since the radiative gluino decay has not been fully investigated at the LHC, we recasted three official ATLAS and CMS analyses within the \texttt{MadAnalysis5} (\texttt{MA5}) framework~\cite{Conte:2012fm,Conte:2014zja} to derive mass limits applicable in a scenario where the gluino loop decay dominates. The recasted searches are: an ATLAS monojet analysis~\cite{Aad:2014nra} and ATLAS and CMS multijets analyses~\cite{Aad:2014wea,Chatrchyan:2014lfa}. Such analyses are now available on the \texttt{MA5} Public Analysis Database~\cite{Dumont:2014tja,MADPAD}. To derive the limits, we defined a SMS scenario where $\mbox{BR}(\tilde g \rightarrow g \widetilde{\chi}_1^0)=100\%$. For the signal, gluino pair production was generated using \texttt{MadGraph5}~\cite{Alwall:2014hca}, then the showering and hadronisation was performed by \texttt{PYTHIA-6.426}~\cite{Sjostrand:2006za}. The results are displayed in Fig.~\ref{fig:8TeVlimits}. More details about the recasting and limiting procedure can be found in~\cite{Chalons:2015vja,MADPAD}. One can see that our interpretation (see the various red lines in Fig.~\ref{fig:8TeVlimits} and in particular the solid red one which corresponds to the recasted ATLAS monojet analysis) of recasted searches is indeed sensitive to cases where the gluino-neutralino mass difference is small and complementary to the official results (green, blue and purple lines in Fig.~\ref{fig:8TeVlimits}).

\section{Conclusion}
The radiative gluino decay can be used as a sensitive probe in scenarios where three-body decays are closed or heavily suppressed. Largest branching fractions for this decay pattern are obtained when the neutralino is higgsino-like. By recasting official analyses within the \texttt{MA5} framework we could exclude in this scenario gluinos degenerate with neutralinos up to $m_{\tilde g} \simeq m_{\widetilde{\chi}_1^0} \simeq 750$ GeV and close the gap above the $t \bar{t}$ threshold, assuming a simplified model where $\mbox{BR}(\tilde g \rightarrow g \widetilde{\chi}_1^0)=100\%$. Closing the small gluino-neutralino mass gap can have important consequences in scenarios where the gluino is the next-to-lightest SUSY particle, in particular in solving the Dark Matter problem using gluino-neutralino coannihilation in the MSSM. The complete investigation~\cite{Chalons:2015vja} of the work summarised here also explored the discovery prospects of the gluino decay through its radiative decay at Run II of the LHC. In particular we showed that, at the Monte Carlo level, a dijet search strategy may be more sensitive than a monojet one.
\section*{Acknowledgments}
The work of GC is supported by the Theory-LHC-France initiative of CNRS/IN2P3. The work of DS is supported by the 
French ANR, project DMAstroLHC, ANR-12-BS05-0006, and by the Investissements d'avenir, Labex ENIGMASS.

\end{document}